# Two approaches to the cosmic ray "knee" problem and the experimental data

Yu. V. Stenkin

*Institute for Nuclear Research of Russian Academy of Sciences*
*60th October anniv. prospect 7a, 117312 Moscow, RUSSIA*
Presenter: Yu.V.Stenkin  (stenkin@sci.lebedev.ru), rus-stenkin-Y-abs1-he12-oral

Two approaches to the old problem of the so-called "knee" in cosmic ray spectrum are compared with the existing experimental data. It is shown that "standard astrophysical model" of the knee is not supported by some experimental data, while the alternative model agrees better with the data.

## 1. Introduction

The problem of the "knee" observed in cosmic ray spectrum in PeV region is one of the crucial points in modern cosmic ray physics and astrophysics. Unfortunately, the intensity of cosmic rays in this region is very low ($\sim 1$ m$^{-2}$ y$^{-1}$) and physicists are pressed to use indirect methods at energies above 1 PeV. The method of Extensive Air Shower (EAS) elaborated more than 50 years ago gave an instrument to study cosmic ray spectrum up to the highest possible energies. But, a shadow side of the method is a very difficult interpretation of the observed data. The Earth's atmosphere is a rather thick calorimeter ($\sim 11$ hadronic interaction lengths) which absorbs almost full energy of the primary particle. Another factor which affects the result uncertainty is unknown mass of primary particle which are spread mostly between A=1 for primary proton and A=56 for primary iron. Big fluctuations in EAS development caused by rather small amount of cascading hadrons in the EAS core produce additional difficulties in the data interpretation. I could remind that the phenomenology of EAS as a hadronic cascade in the atmosphere where secondary electromagnetic component produced through decays of neutral mesons is in equilibrium with the hadronic EAS skeleton, was developed only at the end of 40-s - beginning of 50-s [1] just before the appearance of the work [2] claimed the "knee" existence. I would also remind the readers that at those times recalculation from the EAS size at maximum of cascade curve ($N_e^{max}$) to primary energy ($E_0$) was very simple: $E_0 = \mathbf{k} \cdot N_e^{max}$, where $\mathbf{k}$ is a constant. This means that any change of observed slope in $N_e$ distribution was automatically assigned to primary spectrum.

Widely used astrophysical approach to the knee problem which we called as standard model tries to solves the problem as follows: i) the knee does exist in primary spectrum and is caused by astrophysical reasons, ii) at $E_0 \approx 3$-5 PeV spectrum of primary protons becomes steeper, iii) the knee energy is proportional to the charge of primary nucleus Z (proportional to mass A in some models) and therefore, primaries mass composition becomes heavier above the knee.

An alternative approach to the knee problem (phenomenological approach) was developed in 2003 [3]. It solves the problem as follows: i) primary spectrum follows power law ($\gamma$=const) and primary mass composition is also constant, ii) visible knee in the EAS size spectrum is caused by the break of equilibrium between high energy hadronic and secondary electromagnetic components in the point where the last hadron lost its energy and/or decayed, iii) this energy is equal to $\sim 100$ TeV/nucleon at sea level and below this point EAS has no core (coreless EAS), iv) properties of the coreless EAS' differ significantly from those of normal EAS': it looks like a pure electromagnetic EAS with addition of muons ("e-poor"), v) attenuation of such a shower follows a pure electromagnetic scenario ($\Lambda = \Lambda_{em} \approx 100$ g/cm$^2$) and the latter results in a change of slope $\alpha$ in ratio $N_e \sim E_0^{\alpha}$, vi) the latter means that observed EAS size distribution $I(N_e) \sim N_e^{-\gamma/\alpha}$ should also change a slope in 2 points: at $\sim 100$ Tev – "proton knee" and at $\sim 5$ TeV – "iron knee".

This model thus naturally predicts the appearance of 2 "knees" (only in electromagnetic component), their energy and even absolute value of the visible spectrum slope change. It also predicts the absence of the "knee" in muonic and hadronic components. All these parameters depend on the experimental details (altitude, array size, trigger conditions etc.) and should be calculated using one of the modern Monte Carlo programs. Note that any fit of calculation data for $N_e(E_0) \sim E_0^\alpha$ with constant parameter $\alpha$ is inadmissible as it implies the "knee" in primary spectrum *a priori*.

## 2. Overview of the experimental data

It is impossible to overview a huge number of experiments made during last 50 years in a frame of the Conference paper therefore, I included in the Table 1 below only experiments published mostly during the last decade or presented at recent ICRCs. To save the paper volume I sometimes refer to rapporteur's talks or to reviews and I do not pretend to make complete data overview. I concentrated mostly on the parameters that allow comparing two different approaches on a base of different predictions. It should be noted that in the references I give here, the authors did not pay attention to some features (for example to an existence of the first knee at ~100 TeV) of their data that I found *a posteriori*.

**Table 1**. Comparison with experimental data.

| parameter | Astrophysical approach (standard model) | Phenomenological approach [3] | Experimental data |
|---|---|---|---|
| Age as a function of $E_0$ (below 5 PeV), s1 | Small decrease due to $X_{max}$ movement | Decrease due to appearance of young coreful EAS' | Decrease observed [4, 5] |
| Age as a function of $E_0$ (above 5 PeV), s2 | Rising due to change of mass composition | Constant or small decrease due to $X_{max}$ movement | Constant [4, 5] |
| Position of the "proton" knee at sea level, $N_e^{knee}$ | ~$10^{6.0-6.2}$ | ~$10^{4.8}$ | both observed by [6, 7, 8, 9, 10] |
| Position of the "iron" knee at sea level, $N_e^{knee}$ | ~$10^{7.4-7.6}$ | ~$10^{6.2-6.4}$ | Only at $N_e^{knee}$ ~$10^{6.2}$ observed |
| Position of the "iron" knee at 4-5 Km a.s.l, $N_e^{knee}$ | ~$10^{8.6}$ | ~$10^{6.0}$ | $N_e^{knee}$ ~$10^{6.0}$ [11, 12, 13] |
| Intensity at the knee (5PeV) at sea level, $Log(I_k)$ | -12.75 | $\leq$(-12.75) | -12.75 [11] |
| Intensity at the knee (5PeV) at the altitude 4-5 Km a.s.l., $Log(I_k)$ | -12.75 | ~(-11) | -11.5 [11, 12, 13] |
| Change of EAS size spectrum slope at $N_e$~$10^{4.8-5.0}$, $\Delta\beta_e$ | 0 | ~0.35 for standard composition; ~0.75 for pure proton composition 0 for pure iron composition | ~0.5 [6] ~0.45 [7] 0.5 [9] |
| Change of EAS size spectrum slope at $N_e$~$10^{6.0-6.2}$, $\Delta\beta_e$ | No prediction | ~0.5 for standard composition ~0.75 for pure iron composition 0 for pure proton composition | 0.35÷0.57 |
| Change of EAS size spectrum slope for muons and hadrons, $\Delta\beta_\mu$; $\Delta\beta_h$ | $\Delta\beta_\mu > \Delta\beta_e$ $\Delta\beta_h > \Delta\beta_e$ (see [13]) | ~0 | ~0 [14, 15] $\Delta\beta_\mu \leq 0.2 < \Delta\beta_e$ [16] |

| parameter | Astrophysical approach (standard model) | Phenomenological approach [3] | Experimental data |
|---|---|---|---|
| Difference between old and young EAS' | It is supposed that all young EAS' are originated from p- and all old EAS' from Fe-induced EAS' | Explained by different behavior of *coreless* and *coreful* EAS'. Only *coreful* (young) showers have a knee in PeV region. The *coreless* branch have a knee somewhat below 100 TeV | Change to heavier primaries disagrees with observed *s* decreasing. Behavior of EAS' selected as "very old" and selected as "very young" coincides with that for *coreless* and *coreful* EAS' [3, 4] |
| $X_{max}(E_0)$ in PeV region | Slope changes at $E_0 > 3\text{-}5$ PeV | No changes of slope | Very big dispersion of data: nonstatistical errors [17] |
| $<Ln(A)>$ as a function of $E_0$ | Rise above 3-5 PeV | No changes in mass composition. Visible increase at $5\text{PeV} > E_0 > 100\text{TeV}$ due to rising contribution of heavier primaries to recorded EAS'. Normal composition up to the highest $E_0$ | Nonstatistical spread of data (evidence for systematic errors) but the rise can be seen from $E_0 > 100\text{TeV}$ [17, 19, 20] |
| $<Ln(A)>$ at $E_0 > 1$ EeV | No predictions for highest $E_0$ | | normal proton content above 10 EeV [18] |
| Attenuation length as a function of $E_0$ $\Lambda_{att}$, g/cm$^2$ | Constant | Increase from ~100 at $E_0 < 100$ TeV (when all EAS' are *coreless*) to ~200 at $E_0 > 5$ PeV (when all EAS' are *coreful*) | Increase from ~170 and up to ~200 is observed in the range of $E_0$ ~0.1÷5.0 PeV [20, 21]. |
| Elongation rate (ER) as a function of $E_0$ | Change at $E_0 > 5$ PeV | constant | $X_{max}(E_0)$ can be fitted by a function with constant ER in whole range from $10^{10}$ to $10^{20}$ eV [23; 24] |

## 3. Conclusion

Looking at the Table one could conclude:
- existing experimental data contradict each other and do not agree in many points with the predictions and expectations made on a base of the standard model (it has been shown very well by Prof. Schatz [11]);
- in some parameters such as intensity at the knee, absence of second knee at $E_0 \sim 100$ PeV ("iron knee"), very small visible "knee" in muonic and hadronic components there exist a prompt contradiction with the standard model and some additional assumptions have to be done to remove the contradictions;
- phenomenological model of the knee agrees better with experimental data on many parameters;
- non-statistical spread of experimental data on spectrum slope before and after the "knee", on mass composition etc. could be explained in frames of the phenomenological model by inadequate recovering of primary spectrum parameters when the "knee" existence is accepted *a priori* without any doubt.

Finally I would add that a theory of cosmic ray acceleration with a pure power law spectrum and isotropic sources distribution, does exist [25]. In contrast to other theories it predicts even the integral spectrum slope $\gamma = \sqrt{3} \approx 1.73$, which agrees well with direct spectrum measurements. The EAS method should work properly at energy above ~10 PeV. That is probably why recovered primary spectrum obtained with this method has again the slope very close to 1.73 above ~1 EeV. Moreover, even in absolute units it coincides with the direct measurements data best fit extrapolated from low energies. The latter would be impossible in a case of existence of different cosmic ray sources and different acceleration processes at low and at high energy.

## 4. Acknowledgements

I'd like to express my deep gratitude to Prof. G.T.Zatsepin for valuable contribution to this problem study, for many fruitful discussions and for continuous support of this work.
Author is also grateful to G.V. Domogatsky for useful remarks and support.

This work was supported in part by the RFBR grant # 05-02-17395 and by the Scientific School Program grant 1828.2003.02.